\documentclass[12pt]{article}
\usepackage{amssymb,eufrak}
\begin{document}

\begin{titlepage}
                            \begin{center}
                            \vspace*{0cm}
        \large\bf The geodesic rule for higher codimensional global defects\\

                            \vfill

\normalsize\sc Anthony  J.  Creaco$^\ast$, \ \ \ \ \ Nikos  Kalogeropoulos$^\dagger$\\

                            \vspace{0.2cm}

 \small\sf{ Department of Science,\\
            BMCC-The City University of New York\\
            199 Chambers St., New York, NY 10007, USA}
                            \end{center}

                            \vfill

                     \centerline{\normalsize\bf Abstract}
                            \vspace{0.3cm}
\normalsize\rm\setlength{\baselineskip}{18pt}

We generalize the geodesic rule to the case of formation of higher
codimensional global  defects. Relying on energetic arguments, we
argue that, for such defects, the geometric structures of interest
are the totally geodesic submanifolds. On the other hand,
stochastic arguments lead to a diffusion equation approach, from
which the geodesic rule is deduced. It turns out that the most
appropriate geometric structure that one should consider is the
convex  hull of the values of the order parameter on the causal
volumes whose collision gives rise to the defect. We explain why
these two approaches lead to similar results when calculating the
density of global defects by using a theorem of Cheeger and
Gromoll. We present a computation of the probability of formation
of strings/vortices in the case of a system, such as nematic liquid crystals,
 whose vacuum is \ $\mathbb{R}P^2$.\\

                             \vfill

\noindent\sf PACS: \ \ \ \ \ \ 02.50.Ey, 05.10.Gg, 11.15.Ex \\
Keywords: \ Topological defects, Geodesic rule, Convex Sets, Totally geodesic submanifolds.\\

                             \vfill

\noindent\rule{11.5cm}{0.2mm}\\
\small\rm{$\ast$ \ E-mail: \ acreaco@bmcc.cuny.edu\\
$\dagger$ \  E-mail: \ nkalogeropoulos@bmcc.cuny.edu, \ \
nkaloger@yahoo.com}
\end{titlepage}

%%%%%%%%%%%%%%%%%%%%%%%%%%%%%%%%%%%%%%%%%%%%%%%%%%%%%%%%%%%%%%%%%%%%%%%%%%%%%%%%%

                                 \newpage

\normalsize\setlength{\baselineskip}{18pt}

                        \centerline{\sc 1. \ Introduction}

                                 \vspace{5mm}

Topological defects [1] are static solutions of the Euler-Lagrange
equations of a classical field theory, whose non-triviality is
guaranteed by the homotopy type of the vacuum. The stability of a
topological defect can be ascribed to the infinite amount of
energy required to reduce it to the trivial configuration, whose
energy is zero [1]. This highlights one of the reasons for the
methodological importance of topological defects in field
theories: they provide a glimpse of non-perturbative aspects of
the vacuum of a model, a vacuum for which very few things are
generically known.\\

The geodesic rule [2]-[11] is a way that allows us, in principle,
to make predictions about the density of such topological defects.
There seems to be a significant body of experimental [12]-[16] and
numerical results [17],[18] that makes us be reasonably confident
about the validity of the geodesic rule for the case of global
defects. The formulation and the validity of this rule for local
defects, however, is a much more contentious issue, both
experimentally and numerically [19],[20]. For this reason so we
will confine our analysis to just global defects in the present
work. It is worth noticing that the geodesic rule is not the only
known way for producing topological defects. Another way is
through order parameter oscillations [21],[22],[23] whose
contributions are dominant is
when there is no significant dissipation during bubble collisions. \\

To be more concrete, we will assume that the phase transition
after which a topological defect is formed is of first order and
proceeds by bubble nucleation [1],[10],[11]. Then the geodesic
rule states that, when two such bubbles collide, the order
parameter (spacetime scalar) \ $\phi$ \ smoothly interpolates
between its values in the bubbles \ $\phi_1$, \ $\phi_2$ \
following the shortest path (geodesic) on the vacuum manifold.
There are two ways to justify the validity of geodesic rule: The
first relies on minimizing the energy density of the model under
consideration in the spontaneously broken phase [4],[5]. The
second relies on the stochastic nature of the bubble collisions in
the broken phase and the Markovian character of the bubble
effective dynamics during their expansions and their collisions
[24],[25]. In this paper we extend the arguments of both of these
approaches to the case of formation of higher co-dimensional
defects (monopoles, textures etc). We justify the reason why these
two approaches lead to the same estimate for the density of global
defects by using a theorem of Cheeger and Gromoll [26].  We use an
integral geometric argument to determine the probability of
formation of global strings/vortices formed in a system whose
vacuum is \ $\mathbb{R}P^2$ \ as was also done in an earlier work.\\

                         \vspace{3mm}

%%%%%%%%%%%%%%%%%%%%%%%%%%%%%%%%%%%%%%%%%%%%%%%%%%%%%%%%%%%%%%%%%%%%%%%%

    \centerline{\sc 2. \ The energetic approach and totally geodesic submanifolds}

                        \vspace{5mm}

A classical, quantum or thermal field theory describing the phase
transition that gives rise to topological defects, has as basic
kinematic variables the order parameter(s) \ $\phi$, \  which, for
simplicity, we will assume to be spacetime scalar fields in this
work. Let the spacetime, on which the model is defined, be \
$\mathbb{R}^k$, \ endowed with the usual flat metric, for
concreteness.  The scalar fields \ $\phi$ \ are sections of a
vector bundle with base \ $\mathbb{R}^k$ \ and typical fiber an,
irreducuble most of the times, representation of a compact,
semisimple Lie group \ $G$, \ which incorporates the internal
symmetries of the system. The vacuum then consists of all such
maps \ $\phi$ \ that minimize an appropriate energy functional. In
most cases we choose to focus our attention on the image of the
space-time points under \ $\phi$, \ rather than on the set of such
maps themselves, as the basic kinematic variables. Such a choice
is largely unimportant however, especially within the domain of
perturbation theory around a smooth  \ $\phi$, like the ones used
in the Ginzburg-Landau description of phase transitions, which we
will be working with
in what follows.\\

We assume that the vacuum of the model is an $m$-dimensional
Riemannian manifold \ $\mathcal{M}$, \ endowed with a metric \
$\tilde{g}$. \ In most cases of interest, as was mentioned above,
\ $\mathcal{M}$ \ is a semisimple, compact Lie group \ $G$, \
endowed with its Cartan-Killing metric \ $ \tilde{g} =
Tr(u^{-1}du)^2, \ \ u\in G $. \  Then the Cartan-Killing metric is
non-degenerate and positive definite, so \ $(G, \tilde{g})$ \ is,
indeed, a Riemannian manifold. The vacuum \ $\mathcal{M}$ \ is
also very frequently a homogeneous space \ $G/H$, \ endowed with
the induced metric from $\tilde{g}$. \  The energy functional of
the model can be written as
\begin{equation}
    E [\phi ] \ = \ \int_{\mathbb{R}^k} \{\frac{1}{2} (\partial_{\alpha}\phi)
    (\partial^{\alpha}\phi) + V(\phi) \}
\end{equation}
where \ $\alpha = 1, 2, \ldots, k$ \ and \ $V(\phi)$ \ denotes the
potential energy density of the model. The possible existence of a
mass term \ $\frac{m^2}{2}\phi^2$ \ is assumed to be incorporated
inside \ $V(\phi)$. \ Local energy extremization, under an
infinitesimal energy variation \ $\delta E$ \  gives
\begin{equation}
   \delta E[\phi] \ = \ \int_{\mathbb{R}^k} \{
   (\partial^{\alpha}\phi) \delta(\partial_{\alpha}\phi) + \delta
   V (\phi) \}
\end{equation}
All the points of the stable vacuum have the same potential energy
density, so we can set \ $\delta V(\phi) = 0$ \ on \
$\mathcal{M}$. \ Then the vanishing of the first variation of the
energy reduces to
\begin{equation}
          \partial_{\alpha}\delta\phi = 0
\end{equation}
This implies that the energy between two points of the vacuum \
$\phi_1$ \ and \ $\phi_2$ \ will be extremized when \ $\delta\phi$
\ is constant [4],[5]. This constant can be made equal to zero by
an affine reparametrization. Such a result can be heuristically
interpreted, as stating that the scalar field must follow a path
of ``stationary length", with respect to perturbations, when it is
arc-length parametrized, namely it must follow a geodesic, with
endpoints \ $\phi_1$ \ and \ $\phi_2$. \ Such a geodesic is
locally length minimizing [26],[27], as is also noted right after
eq. (28) below. Generally, the variation \ $\delta$ \ is in the
space of all maps \ $\phi$ \ that are sufficiently smooth for our
purposes. However, since we have decided to focus on the image of
points of \ $\mathbb{R}^k$ \ under \ $\phi$, \ rather than on \
$\phi$ \ themselves, the variations \ $\delta$ \ effectively
reduce to variations on \ $\mathcal{M}$.\\

One can extend, still heuristically, this argument to the case of
higher co-dimensional defects. For concreteness,  let us consider
the case of monopoles, which are produced by the collision of four
causally disconnected bubbles with scalar field values \ $\phi_i,
\ \ i=1, \ldots, 4$. \ According to the geodesic rule, \ $\phi$ \
will interpolate between each pair of bubbles, along the
geodesic(s) \ $\phi_{ij}, i,j = 1, \ldots, 4, \ \ i\neq j$ \ that
join the values of \ $\phi$ \ in these bubbles, on \
$\mathcal{M}$. \ Consider a 2-simplex in \ $\mathcal{M}$ \ with
vertices \ $\phi_i$ \ and \ edges \ $\phi_{ij}$. \ By analogy with
the geodesic rule, and as we will prove in what follows, we should
require the faces of this simplex to be least-area manifolds
having as boundary  the union of the geodesic segments \
$\phi_{ij}$. \ We can then determine the homotopy properties of
the 2-simplex, and decide, in particular, whether it corresponds
to a trivial element of  the second homotopy group \
$\pi_2(\mathcal{M})$. \ If it does, then a monopole will not be
formed, on topological grounds, from this bubble collision.
Otherwise, a monopole formation is possible. By enumerating the
number of simplices with trivial and non-trivial images in \
$\pi_2(\mathcal{M})$ \ with vertices uniformly distributed on \
$\mathcal{M}$, \ we can provide an estimate of the number of
monopoles that will be formed in this phase transition. One can
inductuvely extend this construction to the case of higher
co-dimensional defects.\\

We want to derive a local analytic condition that, if possible, is
satisfied by the heuristic multi-dimensional extension of the
geodesic rule described in the previous paragraph. Such an
extension applies to the formation of $n$-codimensional defects.
The homotopy group guaranteeing the stability of such defects,
when non-trivial, is \ $\pi_{k-n-1}$. \  As an example of this
homotopic classification, when \ $k=3$ \ and \ $n=0$ \ we get the
condition for the stability of monopoles expressed by the
non-trivial \ $\pi_2$. \  Similarly, we get conditions for the
stability of strings/vortices, textures and other defects. Here \
$n$ \ is the dimension of an appropriate sub-manifold \
$\mathcal{N}$ \ properly embedded in $\mathcal{M}$ \ as \ $l:
\mathcal{N} \rightarrow \mathcal{M}$. \ Although \ $l$ \ is a
proper embedding, on physical grounds, i.e. no self-intersections
of \ $\mathcal{N}$ \ should be allowed,it turns out that such a
condition is not too overly restrictive. Indeed, the following
computations can be appropriately generalized when \ $l$ \ are
immersions with finite self-intersections, resulting in integral
$k$-varifolds for instance, without any  major geometric
modifications [27]. We allow for \ $\mathcal{N}$ \ to have a
piecewise smooth boundary \ $\partial\mathcal{N}$. \ This boundary
is composed of simplices iteratively generated by the values of \
$\phi_i, \ i = 1,\ldots, n+2$ \ in the colliding bubbles, the
geodesics $\phi_{ij}, \ i,j = 1, \ldots, n+2, \ i\neq j$ \ joining
them, and their higher dimensional counterparts.  Let \ $g =
l_{\ast}\tilde{g}$ \ indicate the induced metric on \
$\mathcal{N}$ \ and \ $x\in\mathcal{N}$. \ We choose an
orthonormal basis \ ${e_1, \ldots, e_n, \nu_{n+1}, \ldots, \nu_m}$
\ of \ $T_x\mathcal{M}$, \ with \ $e_i, \ i = 1, \ldots, n$ \
orthonormal in \ $T_x\mathcal{N}$ \ and \ $\nu_j, j = n+1, \ldots,
m$ \  othonormal in the fiber \ $N_x\mathcal{N}$. \ Then any \
$X\in T_x\mathcal{M}$ \ can be decomposed as \ $X = X^\top +
X^\bot$ \ where \ $X^\top\in T_x\mathcal{N}$ \ and \ $X^\bot\in
N_x\mathcal{N}$. \ Let \ $\widetilde{\nabla}$ \ and \ $\nabla$ \
denote the Levi-Civita connections on \ $\mathcal{M}$ \ and \
$\mathcal{N}$, \ respectively. Then, for \ $X,Y \in T\mathcal{N}$
[28],
\begin{equation}
    \nabla_X Y = (\widetilde{\nabla}_X Y)^\top
\end{equation}
Consider a smooth isotopy (``smooth deformation") of \
$\mathcal{N}$ \ which keeps  \ $\partial\mathcal{N}$ \ fixed. Such
an isotopy is a map
\begin{equation}
     F(x,t) : \mathcal{N} \times (-\epsilon, \epsilon) \rightarrow
     \mathcal{M}
\end{equation}
with \ $\epsilon > 0$, \ which a smooth enough diffeomorphism \
$F( \cdot, t) : \mathcal{N} \rightarrow \mathcal{N}$, \  for each
$t \in (-\epsilon, \epsilon)$. \ The fact that such a
diffeomorphism exists is guaranteed by the implicit function
theorem. The vector field generating \ $F(x,t)$ \ is
\begin{equation}
           F_t = \frac{\partial F(x,t)}{\partial t}\Bigg|_{t=0}
\end{equation}
and let the induced metric on \ $F(\mathcal{N},t)$ \ be denoted by
\ $g_t$. \ Obviously \ $g_0 = g$. \ According the area formula
[29],
\begin{equation}
    Vol (F(\mathcal{N}, t)) = \int_\mathcal{N} J(x,t) \ d^nx
\end{equation}
where \ $J(x,t)$ \ denotes the Jacobian associated with the action
of \ $F_t$. \ This Jacobian can also be expressed in terms of the
metric as [30]
\begin{equation}
                      J(x,t) = \sqrt{\det g_t}
\end{equation}
which means that
\begin{equation}
   \frac{d}{dt} \ Vol (F(\mathcal{N},t)) \Bigg|_{t=0} =
     \int_\mathcal{N} \left( \frac{d}{dt} \sqrt{\frac{\det g_t}{\det
     g_0}}  \ \Bigg|_{t=0} \right)  \sqrt{\det g_0} \ d^nx
\end{equation}
Since \ $g_t$ \ is a positive-definite symmetric matrix, it
satisfies the identity
\begin{equation}
 \det g_t = \exp(tr \ln g_t)
\end{equation}
where \ $tr$ \ denotes the trace of \ $g_t$ \ and we get
\begin{equation}
   \frac{d}{dt} \ \sqrt{\det g_t} \ \Bigg|_{t=0} = \frac{1}{2} \
   \sqrt{\det g_0} \ \ tr \left( g_0^{-1}\frac{dg_t}{dt}\Bigg|_{t=0}\right)
\end{equation}
which implies, taking into account that $\{ e_i \}, \ i=1, \ldots,
n$ \ is an orthonormal basis, that
\begin{equation}
   \frac{d}{dt} \ \sqrt{\frac{\det g_t}{\det g_0}} \ \Bigg|_{t=0} =
   \frac{1}{2} \ tr \ \frac{dg_t}{dt}\Bigg|_{t=0} =
   \frac{1}{2} \sum_{i=1}^n  \ \frac{d}{dt} \ g_t(e_i, e_i) \ \Bigg|_{t=0}
\end{equation}
where we have used that the operations of \ $tr$ \ and
differentiation of \ $g_t$ \ commute with each other.  Because \
$\widetilde{\nabla}$ \ is a Levi-Civita connection, we have
\begin{equation}
    \frac{d}{dt} \ g_t(e_i, e_i) = 2
    \tilde{g}(\widetilde{\nabla}_{F_t}e_i, e_i)
\end{equation}
and because it is torsion-free, namely
\begin{equation}
   \widetilde{\nabla}_{F_t}e_i - \widetilde{\nabla}_{e_i}F_t =
   [F_t, e_i] = 0
\end{equation}
we find
\begin{equation}
\frac{d}{dt} \ g_t(e_i, e_i) \Bigg|_{t=0} =
\tilde{g}(\widetilde{\nabla}_{e_i}F_t, e_i)
\end{equation}
which, upon using Leibniz's rule with the Levi-Civita condition
once more, gives
\begin{equation}
   \widetilde{\nabla}_{e_i}[\tilde{g}(F_t, e_i)] =
   \tilde{g}(\widetilde{\nabla}_{e_i}F_t, e_i) + \tilde{g}(F_t,
   \widetilde{\nabla}_{e_i}e_i)
\end{equation}
resulting in
\begin{equation}
\frac{d}{dt} \ \sqrt{\frac{\det g_t}{\det g_0}} \ \Bigg|_{t=0} =
     \sum_{i=1}^n \widetilde{\nabla}_{e_i} [\tilde{g}(F_t, e_i)] -
         \tilde{g}(F_t, \nabla_{e_i}e_i)
\end{equation}
Because \ $\tilde{g}(F_t, e_i)$ \ is a scalar field and by the
definition of divergence [27],[30]
\begin{equation}
  div_\mathcal{N} X = \sum_{i=1}^n g(\nabla_{e_i}X, e_i), \ \ \
  \forall \ X\in T_x\mathcal{N}
  \end{equation}
we find that
\begin{equation}
\frac{d}{dt} \ \sqrt{\frac{\det g_t}{\det g_0}} \ \Bigg|_{t=0} =
  div_\mathcal{N}F_t^\top - \sum_{i=1}^n \tilde{g}(F_t,
  \widetilde{\nabla}_{e_i} e_i)
\end{equation}
Combining Stokes' theorem and the initial assumption that \ $F_t =
0$ \ on \ $\partial\mathcal{N}$, \ we conclude that
\begin{equation}
    \int_\mathcal{N} div_\mathcal{N} F_t^\top \sqrt{\det g_0} \ d^nx = 0
\end{equation}
The second term on the right hand side of (19) can be
re-expressed, due to linearity, as
\begin{equation}
\sum_{i=1}^n \tilde{g}(F_t, \widetilde{\nabla}_{e_i}e_i) =
  \sum_{i=1}^n \tilde{g}(F_t, (\widetilde{\nabla}_{e_i}e_i)^\bot)
+ \sum_{i=1}^n \tilde{g}(F_t, (\widetilde{\nabla}_{e_i}e_i)^\top)
\end{equation}
We observe that the second term reduces to
\begin{equation}
  \sum_{i=1}^n \tilde{g}(F_t, (\widetilde{\nabla}_{e_i}e_i)^\top)
  = \sum_{i=1}^n \tilde{g}(F_t, \nabla_{e_i}e_i)
\end{equation}
By choosing a normal coordinate system around \ $x\in\mathcal{N}$,
\ we see that \ $\nabla_{e_i}e_i = 0$ \ and, since this is a
vector relation, it is true in all coordinate systems. Therefore
\begin{equation}
\sum_{i=1}^n \tilde{g}(F_t, (\widetilde{\nabla}_{e_i}e_i)^\top) =
0
\end{equation}
The second fundamental form at \ $x\in\mathcal{N}$ \ is a vector
in \ $N_x\mathcal{N}$ \ given by
\begin{equation}
  S(X,Y) = (\widetilde{\nabla}_X Y)^\bot, \ \ \forall \ X,Y \in
  T_x\mathcal{N}
\end{equation}
We have used the sign convention of [28] in this definition. The
trace of the second fundamental form is called the mean curvature
vector \ $H$, \ so,
\begin{equation}
    H =  \frac{1}{n}\sum_{i=1}^n S(e_i, e_i)
\end{equation}
 Then, the first term of (21) becomes
\begin{equation}
\sum_{i=1}^n \tilde{g}(F_t, (\widetilde{\nabla}_{e_i}e_i)^\bot) =
  n \ \tilde{g}(F_t, H)
\end{equation}
so, (9) eventually gives
\begin{equation}
  \frac{d}{dt} Vol(F(\mathcal{N},t)) \Bigg|_{t=0} =
  \int_\mathcal{N} n \ \tilde{g}(F_t, H) \ \sqrt{det g} \ d^nx
\end{equation}
Since the isotopy \ $F(x,t)$ \ is arbitrary, we conclude from (27)
that the multidimensional generalization of the geodesic rule that
we are seeking, amounts to determining all \ $\mathcal{N}$ \ for
which
\begin{equation}
     H=0,  \ \ \ \ \forall \ x  \in \mathcal{N}
\end{equation}
Submanifolds for which \ $H=0$ \ are called minimal [31],[32].
Obviously, geodesics are minimal 1-dimensional submanifolds of \
$\mathcal{M}$. \ The word ``minimal" may appear to be a misnomer,
since all that we have determined are the extrema of the volume
functional, which are not necessarily minima. This is indeed true
globally. To determine the volume minimizing minimal submanifolds,
one would have to use the second variation formula [28],[29].
Locally, however, a minimal manifold without singularities is
volume minimizing [27]. To be more concrete, assume that \ $x$ \
is not on the boundary \ $\partial\mathcal{N}$. \ It turns out,
then, that there is a small enough open set \
$U\subset\mathcal{N}$, \ with \ $x\in U$ \ such that
\begin{equation}
     Vol (\mathcal{N}\cap U) \leq Vol (F(\mathcal{N}\cap U, t))
\end{equation}
when \ $\mathcal{N}$ \ is minimal. It is worth pointing out that
although an appropriately modified version of (27) is true for
manifolds with singularities, as mentioned above, (29) does not
hold in such a case [27]. An obvious class of embeddings \ $l$ \
satisfying (27) are the ones for which \ $S=0$, \  instead of the
weaker \ $H=0$. \ Such submanifolds \ $\mathcal{N}$ \ are called
totally geodesic [28], because they have the important property
that a geodesic \ $\gamma (s)$ \ of \ $\mathcal{M}$ \ starting at
\ $x \in \mathcal{N}$ \ with an initial direction \ $\frac{d\gamma
(0)}{ds}\in T_x\mathcal{N}$ \ always stays in \ $\mathcal{N}$. \
The order parameter of the broken phase in the colliding bubbles
and the minimal geodesics joining them should be entirely inside \
$\mathcal{N}$. \ In any other case, there is at least a part of a
geodesic segment which is in the complement of \ $\mathcal{N}$, \
thus failing to extremize the energy (1). Such cases are ruled out
by having requiring the extremization (2). This argument
demonstrates that we must choose as physically relevant only the
subset of minimal submanifolds that are totally geodesic. These
are the higher dimensional generalizations of the geodesic rule
that we
were seeking. Obviously the geodesics themselves fulfill this requirement. \\

                                \vspace{3mm}

%%%%%%%%%%%%%%%%%%%%%%%%%%%%%%%%%%%%%%%%%%%%%%%%%%%%%%%%%%%%%%%%%%%%%%%%%%%%%

\centerline{\sc 3. \ Stochasticity, convexity and equivalence of
results}

                                \vspace{5mm}

In previous work [24],[25] one of us presented an argument leading
to the geodesic rule on stochastic grounds. The argument relied,
in an essential way, on the existence of two widely distinct time
scales, one relatively short for the bubble collisions \ $\tau_C$,
\ and a much longer one for the bubble coalescence \ $\tau_M$. \
With the additional assumptions of the Markovian character of the
bubble dynamics and the almost isotropy of the vacuum \
$\mathcal{M}$, \ a diffusion equation governing the evolution of
the order parameter inside the bubbles in the low-temperature
phase was derived. The short ``time" asymptotics of the
corresponding solution gave rise to the geodesic rule
[24],[25].\\

There is no obvious a priori way to generalize this result and
determine the higher dimensional analogues of the geodesics, call
them \ $\mathcal{Q}$, \ in the stochastic approach. However we can
proceed inductively as follows: Let's consider the value of the
order parameter \ $\phi_i, \ i = 1,2,3$ \ inside three bubbles
that collide to potentially give rise to a string-like defect.
Assume that these three values are within the injectivity radii of
each other so that the geodesic rule as argued in [24] holds. Let
\ $\phi_{ij}, \ i,j = 1,2,3, \ i\neq j$ \ denote the minimal
geodesics joining the bubbles whose order parameter is \ $\phi_i$
\ and \ $\phi_j$. \ Due to the proximity of \ $\phi_i$ \ and \
$\phi_j$ \ there is just one such segment (minimal geodesic) for
each pair \ $ij, \ i\neq j$. \ Let \ $p\in \phi_{ij}$ \ and \
$q\in \phi_{ik}$ \ with \ $i\neq k$. \ Then, the unique geodesic \
$\phi_{pq}$ \ must be contained in the set \ $\mathcal{Q}$, \ due
to continuity. Continuity also demands all pairs of interior
points of \ $\mathcal{Q}$ \ be connected by segments. All the
points of such segments must belong to \ $\mathcal{Q}$. \ In
Euclidean space we would call a set $\mathcal{Q}$ \ satisfying
such a requirement ``convex". We must be careful, however, since
there are several inequivalent concepts of convexity in the
Riemannian context, all of which coincide in the Euclidean case
[23],[25],[27]. To state the problem: what we are seeking is a set
\ $\mathcal{Q}\subset\mathcal{M}$, \ such that for any two \ $p,q
\in\mathcal{Q}$ \ there exists a segment \ $\phi_{pq} \subset
\mathcal{Q}$ \ such that \ $\phi_{pq}$ \ is the unique segment
connecting \ $p$ \ and \ $q$ \ in \ $\mathcal{M}$. \ The set \
$\mathcal{Q}$ \ satisfying this condition is called ``convex" even
in the Riemannian setting [26],[28],[30]. The geometric structure
\ $\mathcal{Q}$ \ we wanted to determine, therefore, is the convex
hull  \ of \ $\phi_i, \ i=1,2,3$ \ in \ $\mathcal{M}$. \ We see
that this result can be inductively generalized for all higher
dimensional topological defects, without any further difficulty.\\

The class of classical vacua \ $\mathcal{M}$ \ usually employed
satisfies one further condition. As was pointed out above, \
$\mathcal{M}$ \ is very frequently a Lie group \ $(G,\tilde{g})$,
\ where \ $\tilde{g}$ \ is the Cartan-Killing metric or one of its
homogeneous spaces \ $G/H$ \ with the induced metric. \ For such
cases more can be stated. Indeed, let \ $W,Z,V \in TG$ \ be
left-invariant vector fields on \ $G$ \ and \ $[W,Z]$ \ be their
commutator.  If \ $ |W\wedge Z| = \tilde{g}(W,W) \tilde{g}(Z,Z) -
[\tilde{g}(W,Z)]^2$ \ and the Riemann tensor is
\begin{equation}
 R(W,Z)V = \nabla_W\nabla_Z V - \nabla_Z\nabla_W V - \nabla_{[W,Z]}V
\end{equation}
it turns out that the sectional curvature
\begin{displaymath}
 K(W,Z)= \frac{\tilde{g}(R(W,Z)Z,W)}{|W\wedge Z|}
\end{displaymath}
can be computed to be [28]
\begin{equation}
     K(W,Z) =  \frac{\tilde{g}([W,Z],[W,Z])}{4 |W\wedge Z|}
\end{equation}
Therefore any Lie group \ $(G, \tilde{g})$ \ has non-negative
sectional curvature. The same is true for its homogeneous spaces \
$G/H$ \ with their induced metrics. To see that, it suffices to
notice that the principal fibration \ $H\rightarrow
G\stackrel{\pi}{\rightarrow} G/H$, \ with \ $\pi$ \ being the
natural projection, is a Riemannian submersion. Then, a formula
[33] gives
\begin{equation}
   K(\pi_{\ast}W, \pi_{\ast}Z) = K(W,Z) + \frac{3}{4} \
   \tilde{g}([W,Z]^\bot, [W,Z]^\bot)
\end{equation}
where \ $W,Z$ \ are in addition assumed to be orthonormal and, in
accordance with the notation of Section 2, \ $W^\bot$ \ denotes
the normal component (fiber tangential) of \ $W \in TG$. \
Therefore we are interested in the convexity properties of subsets
\ $\mathcal{Q}$ \ of vacua \ $\mathcal{M}$ \ which are compact
manifolds of non-negative sectional curvature. Consequently the
Ricci curvature along a particular direction of \ $T\mathcal{M}$,
\ which is the average of the sectional curvatures along all
two-planes containing this direction [28],[30], is strictly
positive. The most important point for the stochastic approach
isn't so much that the Ricci curvature on \ $\mathcal{M}$ \ is
positive, but that it has a lower bound. Such a lower bound
guarantees that the diffusion equation that is used to establish
the geodesic rule on \ $\mathcal{M}$ \ has several ``reasonable",
from a physical viewpoint, properties such as the total
conservation of heat (stoachstic completeness), uniqueness of the
associated heat kernel etc [34]. Such properties can be expected
to be true on general physical grounds even if the underlying
smooth structure of \ $\mathcal{M}$ \ were to be abandoned, a fact
toward which the quantum/thermal cases seem to be pointing, as
will be explained in the next Section. \\

It is of some interest to generically ascertain that such convex \
$\mathcal{Q}$, \ as predicted by the geodesic rule, exist for
physically relevant vacua \ $\mathcal{M}$. \ If this were not
true, then topological defects would not form even when they
should have, despite the homotopic arguments  and the experimental
and numerical evidence to the contrary. Then the stochastic
approach, as we have developed it, would be invalid. As a crude
first attempt, one can at least try to guarantee the existence of
convex subsets of \ $\mathcal{M}$ \ by estimating their maximum
``size". An upper bound for the linear dimensions of such a convex
subset is provided by the convexity radius [30] \
$\mathrm{Conv}(\mathcal{M})$. \ It is defined for \
$x\in\mathcal{M}$ \ by
\begin{equation}
  \mathrm{Conv}_x  \ = \sup  \ \{ \rho : B_x(r) \ \mathrm{is \ convex \
                     for \ all} \ r < \rho \}
\end{equation}
where \ $B_x(r)$ \ stands for the solid ball centered at \ $x$ \
of radius \ $r$. \ Then it can be proved that [35]
\begin{equation}
        \mathrm{Conv}(\mathcal{M}) \leq
        \frac{\mathrm{inj}(\mathcal{M})}{2}
\end{equation}
where \ $\mathrm{inj}(\mathcal{M})$ \ indicates the injectivity
radius, which is the minimum distance around any \
$x\in\mathcal{M}$ \ for which the exponential map is a
diffeomorphism. There are various upper and lower bounds for the
injectivity radius of a manifold, depending on conditions that the
curvature, volume and other geometric characteristics of the
manifold satisfy, bounds that we omit since are not explicitly
needed in the sequel [36]. It is encouraging that the convexity
radius is not identically equal to zero, although this may happen
for some \ $x\in\mathcal{M}$. \ It would be useful to be able to
find a lower bound estimate for the convexity radius. Such an
estimate, however, is more difficult to obtain [36], but careful
examination shows that this is not as serious a problem as it
might appear at a first glance. Indeed, the existence of convex \
$\mathcal{Q}$ \ is guaranteed and explicitly provided by the
interior construction of Cheeger and Gromoll [26] that leads to
the proof of their structure theorem mentioned in the next paragraph.\\

As we have seen, the generalizations of the geodesic rule
resulting from the energetic and the stochastic approaches are
embedded totally geodesic submanifolds and  convex subsets of \
$\mathcal{M}$, \ respectively. The question that naturally arises
is whether these two classes of objects give rise to the same
prediction about the density of topological defects. We expect
that they should, since they describe the same physical phenomenon
from two different viewpoints. This expectation turns out to be
correct, as proved by Cheeger and Gromoll in the fundamental [26].
In that work, it was demonstrated that if \ $\mathcal{Q}$ \ is a
closed connected convex subset of a Riemannian manifold \
$\mathcal{M}$, \ then \ $\mathcal{Q}$ \ has the structure of an
embedded $n$-dimensional submanifold of \ $\mathcal{M}$ \ with a
smooth totally geodesic interior and possibly non-smooth boundary
\ $\partial\mathcal{Q}$. Comparing the properties of \
$\mathcal{N}$ \ in Section 2 and of \ $\mathcal{Q}$ \ of the
present Section, we see that the Cheeger-Gromoll theorem implies
that for each \ $\mathcal{Q}$ \ there is exactly one admissible \
$\mathcal{N}$. \ Therefore, the predictions about the density of
topological defects derived by the energetic and by the stochastic
arguments coincide, as they should. The proof of the
Cheeger-Gromoll theorem is attained in several steps and it is
quite geometric. The methods employed as well as the constructions
leading to the result, although highly transparent, appear to have
only limited utility for physical
purposes, so we skip them and refer to the original [26] for further details.\\

                               \vspace{3mm}

%%%%%%%%%%%%%%%%%%%%%%%%%%%%%%%%%%%%%%%%%%%%%%%%%%%%%%%%%%%%%%%%%%%%%%%%%%%%

\centerline{\sc 4. \ Further comments and a sample calculation}

                               \vspace{5mm}

In this Section we are making three mutually loosely connected
comments, related to the topics discussed above. First, we comment
on the time scale in which the geodesic rule was established. As
mentioned in Section 3, the stochastic derivation of the geodesic
rule relied on the assumption that \ $\tau_M \gg \tau_C$ \
[24],[25]. In the case \ $\tau_M \sim \tau_C$, \ the strong mixing
of the order parameters of the colliding bubbles suppresses the
number of topological defects that can be formed. This can occur,
for instance, when there is a high rate of bubble nucleation. To
enforce this suppression explicitly, we can introduce as a
regulator a ``mass" parameter \ $M$ \ in the diffusion equation
that established the geodesic rule. Then the diffusion equation
would become
\begin{equation}
\frac{\partial\phi}{\partial t} = (D + M^2){\nabla}^2 \phi
\end{equation}
where \ $D$ \ is an effective diffusion constant characterizing
the dynamics of \ $\phi$ \ in the broken symmetry phase carried by
the colliding bubbles. The exact form of such a regulating mass
term \ $M$ \ should, ideally, be derived from the dynamics of
bubble collisions. However it is sufficient for our purposes to
know that \ $M\sim 0$ \ for \ $t\sim\tau_C$ \ and that \
$M\rightarrow\infty$ \ for \ $t\sim\tau_M$. \ This behavior can
easily be enforced by the ``time"-dependence
\begin{equation}
M=M_o t^\sigma
\end{equation}
where \ $M_o>0, \ \sigma>0$ \ are constants. The effect of the
addition of \ $M$ \ in to the diffusion equation can be clearly
seen in the, simple and instructive, case of \ $\mathcal{M} =
\mathbb{R}^m$. \ Then the ``massive" heat kernel \ $K_M(\phi_1,
\phi_2; t)$ \ gets expressed in terms of the ``massless" heat
kernel \ $K_0(\phi_1, \phi_2; t)$ \ by
\begin{equation}
      K_M(\phi_1, \phi_2; t) = K_0(\phi_1, \phi_2; t) \ e^{-tM^2D}
\end{equation}
Combining (36) and (37), we observe that the presence of \ $M$ \
suppresses super-exponentially the contributions of the large \
$t$ \ to the heat kernel, which is the desired result. The basic
role of (37) rests in the fact that the ``massive" heat kernel \
$K_M(\phi_1, \phi_2; t)_\mathcal{M}$ \ for any generic vacuum
manifold \ $\mathcal{M}$ \ is expressed as an asymptotic expansion
of \ $K_M(\phi_1, \phi_2; t)_{\mathbb{R}^k}$ \ in terms of \ $t$ \
through [34]
\begin{equation}
  K_M(\phi_1, \phi_2; t)_\mathcal{M} = K_M(\phi_1, \phi_2; t)_{\mathbb{R}^k}
        \left\{ 1 + t a_2(\phi_1, \phi_2) + t^2 a_4(\phi_1, \phi_2) +
        \ldots \right\}
\end{equation}
Here \ $a_2, a_4, \ldots $ \ are polynomials of the Riemann tensor
of \ $\mathcal{M}$, its contractions, its covariant derivatives,
and subsequent combinations of them [37]. We see that, due to (38)
the super-exponential decay of the large \ $t$, \ contributions
still persist for any \ $\mathcal{M}$ \ as in the case of \
$\mathbb{R}^m$. \ The effect of suppressing  the large \ $t$ \
contribution through \ $M$ \ can also be seen at the level of the
scalar propagator on \ $\mathcal{M}$ \ which is [37]
\begin{equation}
       Q^{-1}(\phi_1, \phi_2)_\mathcal{M} = \int_0^\infty
       K_M(\phi_1,\phi_2;t)_\mathcal{M} \ dt
\end{equation}
which when combined with (37) gives
\begin{equation}
  Q^{-1}(\phi_1, \phi_2)_\mathcal{M} = \int_0^\infty
       K_0(\phi_1,\phi_2;t)_\mathcal{M} \ e^{-tM^2D} \ dt
\end{equation}

Second, we have to notice that everything we have mentioned so far
applies for classical or zero-temperature field theories. But it
is far more realistic, and thus desirable, for this formulation to
be applicable when quantum or thermal contributions are taken into
account. Although these two classes of contributions are quite
distinct physically, they are handled by very similar methods,
within the regime of perturbation theory. For this reason from now
on, we will refer only to quantum corrections. Quantum corrections
can be taken into account by replacing the potential energy
density \ $V$ \ by the effective potential \ $V_{eff}$ \ [38] in
(1). To illustrate our points we rely, for simplicity, on one of
the most studied models, the \ $\phi^4$ \ model, whose classical
Lagrangian is, for \ $\alpha = 1,2,\ldots,k$
\begin{equation}
L(\phi) = \frac{1}{2}(\partial^\alpha\phi)(\partial_\alpha\phi) +
          \frac{1}{2} \tilde{m}^2\phi^2 + \frac{\lambda}{4!}\phi^4
\end{equation}
Its quantum effective potential \ $V_{eff}$ \ has a loop expansion
given by
\begin{equation}
     V_{eff} = V_{cl}+ \hbar V_1 + \hbar^2 V_2 + \ldots
\end{equation}
where \ $V_{cl} = \frac{\tilde{m}^2}{2}\phi^2 +
\frac{\lambda}{4!}\phi^4$. \ The one-loop corrections to \
$V_{eff}$ \ receive contributions from the following two mometum
integrals, for \ $\epsilon>0$
\begin{equation}
   \frac{\lambda\phi^2}{4!} \int \frac{d^kp}{(2\pi)^k}
         \frac{i}{p^2-\tilde{m}^2+i\epsilon}  \hspace{1cm} \mathrm{and}
         \hspace{1cm}
   \frac{i}{4}\left(\frac{\lambda\phi^2}{2}\right)^2
        \int \frac{d^kp}{(2\pi)^k}\frac{1}{(p^2-\tilde{m}^2+i\epsilon)^2}
\end{equation}
for the propagator and the four-vertex diagram, respectively [38].
These integrals show explicitly how the dimension \ $k$ \ of
spacetime \ $\mathbb{R}^k$ \ enters the effective potential. Since
the vacuum minimizes \ $V_{eff}$, \ the quantum corrections,
unlike the classical potential, bring an explicit dependence of \
$V_{eff}$ \ on \ $k$. \ This behavior is also in accordance with
the experimental data: the density of topological defects is
observed to depend on the spacetime dimension [1]. In the
physically important case \ $k=4$ \ if we carry out explicitly
these integrals we find that [38]
\begin{equation}
   V_{eff} = \frac{\tilde{m}^2\phi^2}{2} + \frac{\lambda\phi^4}{4!} +
   \frac{1}{(8\pi)^2}\left\{ \left(\frac{\lambda\phi^2}{2} +
   \tilde{m}^2\right)^2 \ln \left(1+\frac{\lambda\phi^2}{2\tilde{m}^2}\right)
   - \frac{\lambda\phi^2}{2}\left( \frac{3\lambda\phi^2}{4} +
   \tilde{m}^2\right)\right\}
\end{equation}
We observe that this expression has a divergence for \
$\tilde{m}=0$. \ This is due to the ultraviolet subtractions in
the one loop contribution, which were set up to enforce the
condition \ $\frac{d^4V_{eff}}{d\phi^4}|_{\phi=0} = 0$. \ In case
we want to extrapolate the calculation of \ $V_{eff}$ \ to \
$\tilde{m}=0$, \ we choose a new renormalization scale \ $\phi =
\tilde{M}$ \ and a renormalized coupling constant \
$\lambda_{\tilde{M}}$ \ at that scale and we obtain the familiar
Coleman-Weinberg expression [39]
\begin{equation}
    V_{eff} = \lambda_{\tilde{M}}\frac{\phi^4}{4!} +
    \frac{\lambda_{\tilde{M}}\phi^4}{(16\pi)^2}
           \left( \ln\frac{\phi^2}{\tilde{M}^2} - \frac{25}{6}
           \right) + \ldots
\end{equation}
We observe that \ $V_{eff}$ \ has a singularity when \
$\ln\frac{\phi^2}{\tilde{M}^2} = \frac{25}{6}$. \ It is very
typical for \ $V_{eff}$ \ to have such singularities. Upon
minimization of \ $V_{eff}$ \ to determine the vacuum, such
singularities will persist, will become more numerous and get more
complicated as the order of the loop expansion increases [38]. As
a result, the vacuum \ $\mathcal{M}$ \ will inherit such
singularities, so when the quantum corrections are taken into
account we can no longer assume \ $\mathcal{M}$ \ to be a
manifold. Although the exact geometric role of such singularities
is unknown, it is clear that the underlying differentiable
structure of \ $\mathcal{M}$ \ loses its smoothness on them. Since
higher orders of perturbation theory keep bringing more and more
such singularities in the determination of \ $\mathcal{M}$, \ we
wonder whether it makes any sense to speak about a smooth
structure of \ $\mathcal{M}$ \ at all ! One way out of such a
difficulty would be to excise, by hand, all the singularities of \
$\mathcal{M}$. \ This is an iterative and ad hoc process, however,
so it is less than satisfying, theoretically. Another approach,
would be to enlarge the structure that \ $\mathcal{M}$ \ can be
allowed to have, from that of  a manifold to that of a metric
space. The advantage of this enlargement is that ``mild"
singularities are already incorporated in the formalism of metric
spaces [27],[29]. Moreover, there is no a priori requirement of
smoothness of such a \ $\mathcal{M}$. \ In this context, it
appears that the metric and measure structures of \ $\mathcal{M}$,
\ which now, unlike the Riemannian case are disjoint, will keep
existing with appropriate modifications, despite the presence of
singularities. Therefore, when quantum corrections are taken into
account we may have to abandon the Riemannian structure of \
$\mathcal{M}$ \ in favor of a metric-measure space structure
[40],[42]. In such a case one can still define a diffusion
equation like the one that gave rise to the geodesic rule, even
without any smoothness assumptions [41],[42]. We find it very
unlikely that in such a case the vacuum will maintain its positive
sectional curvature (now defined in the sense of
Alexandrov-Toponogov) [42], as in the classical/non-thermal case.
However, we  expect that relative volume increments (as defined
through a Hausdorff measure) of \ $\mathcal{M}$ \ would have an
upper bound, a fact which, in the Riemannian case, corresponds to
a lower bound on the Ricci curvature [30]. Lower Ricci curvature
bounds are necessary for the stochastic completeness of the
diffusion equation in the Riemannian case [34] and this may still
turn out to be true for the metric-measure spaces of interest
[42]. \\

Last, we provide a sample of an analytic computation regarding the
probability of formation of vortices (string-like defects) in a
system whose vacuum is \ $\mathbb{R}P^2$. \ An example of such a
system is furnished by nematic liquid crystals.  This calculation
can also be found in [43]. We view \ $\mathbb{R}P^2$ \ as the unit
sphere \ $\mathbb{S}^2$ \ with opposite points identified. Such a
vacuum can arise as a result of the symmetry breaking pattern \
$SO(3) \rightarrow SO(2)\times\mathbb{Z}_2$, \ for instance. To
begin with, it is possible to have vortex formation in this case,
since \ $\pi_1(\mathbb{R}P^2) = \mathbb{Z}_2$. \ To compute this
probability, we have to assume that a vortex is formed from the
collision of three bubbles, the order parameters of which \
$\phi_i, \ i=1,2,3$ \ are independent random variables uniformly
distributed, with respect to the Lebesgue measure, namely the
area, on \ $\mathbb{R}P^2$. \ Without loss of generality, we can
take \ $\phi_1$ \ located at the north pole of \ $\mathbb{S}^2$, \
which is the double covering of \ $\mathbb{R}P^2$. \ The
probability we are seeking places the second point \ $\phi_2$ \ at
an angular distance of at least \ $\pi/2$ \ from \ $\phi_1$, \
since we are interested in a vortex formation. This probability is
given by the ratio of the area of the spherical cap around the
north pole ($\phi_1$) of angular radius \ $\theta$ \ over the area
of the northern hemisphere. We consider the northern hemisphere
since a value of \ $\phi_2$ \ inside it does not give rise to a
vortex, according to the geodesic rule and what follows. Such an
area ratio equals \ $\frac{2\theta}{2\pi} = \frac{\theta}{\pi}$, \
for \ $0\leq\theta <\pi$. \ Then \ $\phi_3$ \  can be located
uniformly anywhere in the hemisphere whose equator is the great
circle joining \ $\phi_1$ \ with \ $\phi_2$ \ since \
$\phi_3\in\mathbb{R}P^2$ \ instead of \ $\mathbb{S}^2$. \ By using
spherical coordinates, the probability of formation of a vortex is
given by
\begin{equation}
   \int_0^{\frac{\pi}{2}} \ \frac{\theta}{\pi} \sin\theta \
   d\theta \ = \ \frac{1}{\pi}
\end{equation}
This type of integral geometric computation  is possible
analytically only in very limited cases, in which the geometry is
so simple as to allow a straightforward enough parametrization of
\ $\mathcal{M}$ \ and subsequent explicit integrations. In all
other cases, such as almost all computations of the density of
higher codimensional defects, a numerical estimate seems to be the
only feasible way to reach any results.\\

                           \vspace{2mm}

%%%%%%%%%%%%%%%%%%%%%%%%%%%%%%%%%%%%%%%%%%%%%%%%%%%%%%%%%%%%%%%%%%%%%%%%%%%%%%%

{\sc Acknowledgement:} We would like to thank the referee for a
critical reading of the manuscript, for several suggestions on how
to improve it and for bringing to our attention references [21] and [43].\\

%%%%%%%%%%%%%%%%%%%%%%%%%%%%%%%%%%%%%%%%%%%%%%%%%%%%%%%%%%%%%%%%%%%%%%%%%%%%%%%

                            \vspace{5mm}

                      \centerline{\sc References}

                            \vspace{0mm}

\begin{enumerate}
 \item A. Vilenkin, E.P.S. Shellard, \ ``Strings and Other
 Topological Defects", Cambridge Univ. Press, Cambridge (1994)
 \item T.W.B. Kibble, \ J. Phys. A {\bf 9}, 1387 (1976)
 \item T. Vachaspati, A. Vilenkin, \ Phys. Rev. D {\bf 30}, 2036 (1984)
 \item A.M. Srivastava, \ Phys. Rev. D {\bf 45}, R3304, (1992)
 \item A.M. Srivastava, \ Phys. Rev. D {\bf 46}, 1353 (1992)
 \item W.H.Zurek, \ Phys. Rept. {\bf 276}, 1977 (1996)
 \item L.Pogosian, T. Vachaspati, \ Phys. Lett. B {\bf 423}, 45 (1997)
 \item N.D. Antunes, L.M.A. Bettencourt, M. Hindmarsh, \
        Phys. Rev. Lett. {\bf 80}, 908 (1998)
 \item A. Ferrera, \ Phys. Rev. D {\bf 59}, 123503 (1999)
 \item A. Rajantie, \ Contemp. Phys. {\bf 44}, 485 (2003)
 \item M. Donaire, \ J. Phys. A {\bf 39} 15013 (2006)
 \item I. Chuang, R. Durrer, N. Turok, B. Yurke, \ Science
         {\bf 251}, 1336 (1991)
 \item M. Bowick, L. Chandar, E.A. Schiff, A.M. Srivastava,
         Science {\bf 263}, 943 (1994)
 \item P.C. Hendry et al. \ Nature {\bf 368}, 315 (1994)
 \item V.M.H. Ruutu et al. \ Nature {\bf 382}, 334 (1996)
 \item R. Carmi, E. Polturak, G. Koren, \ Phys. Rev. Lett.
          {\bf 84}, 4966 (2000)
 \item J. Ye, R. Brandenberger, \ Mod. Phys. Lett. A {\bf 5}, 157
       (1990)
 \item A. Srivastava, \ Nucl. Phys. B {\bf 346}, 149 (1990)
 \item S. Rudaz, A.M. Srivastava, \ Mod. Phys. Lett. A {\bf 8}, 1443
                (1993)
 \item M. Hindmanrsh, A.C. Davis, R. Brandenberger, \ Phys. Rev. D
         {\bf 49}, 1944 (1994)
 \item S. Digal, S. Sengupta, A.M. Srivastava, \ Phys. Rev. D
         {\bf 55}, 3824 (1997)
 \item S. Digal, S. Sengupta, A.M. Srivastava, \ Phys. Rev. D
       {\bf 56}, 2035 (1997)
 \item S. Digal, R. Ray, S. Sengupta, A.M. Srivastava, \ Phys.
       Rev. Lett. {\bf 84}, 826 (2000)
 \item N. Kalogeropoulos, \ Int. J. Mod. Phys. A {\bf 21}, 1493 (2006)
 \item N. Kalogeropoulos, \ Mod. Phys. Lett. A {\bf 21}, 1727 (2006)
 \item J. Cheeger, D. Gromoll, \ Ann. Math. {\bf 96}, 413 (1972)
 \item L. Simon, \ in ``Seminar on minimal submanifolds", E.
 Bombieri, Ed., pp. 3-52, Princeton Univ. Press, Princeton, NJ
 (1983)
 \item T. Sakai, \ ``Riemannian Geometry", \ AMS, Providence, RI
 (1996)
 \item H. Federer, \ ``Geometric Measure Theory", \
 Springer-Verlag, New York (1969)
 \item I. Chavel, \ ``Riemannian Geometry", \ Cambridge Univ.
 Press, Cambridge (1993)
 \item J. Simons, \ Ann. Math. {\bf 88}, 62 (1967)
 \item H.B. Lawson Jr. \ ``Lectures on Minimal Submanifolds", \
 Vol.I, \ Publish or Perish, Boston (1980)
 \item B. O'Neill, \ Mich. Math. J. {\bf 13}, 459 (1966)
 \item I. Chavel, \ ``Eigenvalues in Riemannian Geometry", \
 Academic Press, New York (1984)
 \item M. Berger, \ in ``Differential Geometry and Relativity",
   \ M. Cahen, M. Flato, Eds., pp. 33-42, D. Reidel Pub. Co.,
   Dordrecht (1976)
 \item U. Abresch, W. Meyer, \ in ``Comparison Geometry", \
   K. Grove, P. Petersen, Eds., pp. 1-48, Cambridge Univ. Press,
   Cambridge (1997)
 \item D.V. Vassilevich, \ Phys. Rept. {\bf 388}, 279 (2003)
 \item J. Iliopoulos, C. Itzykson, A. Martin, \ Rev. Mod. Phys.
   {\bf 47}, 165 (1975)
 \item S. Coleman,  E. Weinberg, \ Phys. Rev. D {\bf 7}, 1888
 (1973)
 \item J. Heinonen, \ Bull. Amer. Math. Soc. {\bf 44}, 163 (2007)
 \item Yu. Burago, M. Gromov, G. Perel'man, \ Usp. Mat. Nauka
   {\bf 42}, 3 (1992)
 \item J. Lott, \ ``Optimal transport and Ricci curvature for
 metric-measure spaces", \ {\sf arXiv:math/0610154}
 \item T. Vachaspati, \ Phys. Rev. D {\bf 44}, 2723 (1991)
\end{enumerate}

                               \vfill

\end{document}